# Validation of IRDFF-II library in VR-1 reactor field using thin targets

Michal Kostal[1], Evzen Losa[1,2], Martin Schulc[1], Jan Simon[1], Tomas Bily[2], Vojtech Rypar[1], Martin Mareček[1], Jan Uhlíř[1], Tomáš Czakoj[1], Roberto Capote[3], Andrej Trkov[3], Stanislav Simakov[4]

[1] Research Center Rez, 250 68 Husinec-Rez 130, Czech Republic
[2] Dept. of Nuclear Reactors, Faculty of Nuclear Sciences and Physical Engineering, Czech Technical University in Prague, V Holesovickach 2, Prague 180 00, Czech Republic
[3] Nuclear Data Section, International Atomic Energy Agency A-1400 Wien, Austria
[4] Institute for Neutron Physics and Reactor Technology, Karlsruhe Institute of Technology, D-76344 Eggenstein-Leopoldshafen, Germany

Email: Michal.Kostal@cvrez.cz
Telephone: +420266172655

Abstract

Cross section data are fundamental quantities which affect the accuracy of all calculations in nuclear applications. A new dosimetry library IRDFF-II that contains cross section evaluations with full uncertainty quantification was developed by the International Atomic Energy Agency and released in January 2020 (https://www-nds.iea.org/IRDFF). A previous version, IRDFF-1.05, was released in 2014 and experimental validation of the newly released cross section by spectrum averaged cross section (SACS) measurements is a high priority task. For such purpose, a neutron dosimeter set containing 5 target foils was activated in 2 independent experiments at the VR-1 reactor of the Czech Technical University in Prague. Care was taken to derive SACS with low uncertainties. New experimental evaluation method is in good agreement with previous approaches based on relative measurements using monitor foils. Good agreement of measured SACS and evaluated IRDFF-II cross sections is observed. Slight overestimation of evaluated ENDF/B-VIII.0 $^{235}U(n_{th},f)$ PFNS above 10 MeV is discussed.

## 1 Introduction

The validation of the dosimetry libraries has been traditionally undertaken by measuring spectrum averaged cross section (SACS) in a well-characterized neutron spectrum. Besides the $^{252}Cf(sf)$ prompt fission neutron spectrum (PFNS) standard, recently the ENDF/B-VIII.0 (Brown et al., 2018) $^{235}U(n_{th},f)$ PFNS (Capote et al., 2016, Trkov and Capote, 2015, Trkov et al., 2015) was also declared a reference spectrum (Carlson et al., 2018). Validation of SACS can be undertaken using the dosimetry library evaluations and results compared with measured data. On the other side, if we trust recommended SACS we can judge the quality of the evaluated PFNS.

SACS validation using integral quantities is very useful because the SACS can be measured with significantly lower uncertainties than those propagated uncertainties from differential nuclear data and reference neutron spectra. There are different ways of measuring SACS for high-threshold reactions in a well-defined neutron spectrum. Thanks to non-negligible reactor power used in these experiments, approximately 500 W, the large neutron flux allowed using small activation targets. A typical activation stack was used where dosimetry foils were sandwiched between the monitoring foils. This setup ensures that the neutron flux in the monitoring foils is approximately the same as the neutron flux in the targeted dosimetry foils. Evaluation using normalization to the SACS of monitoring foils is thus enabled. For monitoring, $^{58}$Ni(n,p) reaction is often used (Steinnes E., 1970, Arribére et al., 2001) or combination of more reactions like $^{58}$Ni(n,p), $^{27}$Al(n,$\alpha$) and $^{54}$Fe(n,p) in Maidana et al., 1994, or $^{115}$In(n,n'), $^{58}$Ni(n,p), $^{27}$Al(n,$\alpha$) in Kobayashi et al., 1976.

However, thanks to the existence of validated physical model of the VR-1 reactor, the comparison of the classical and new approaches developed at the zero power LR-0 reactor was tested as well. Being the LR-0 reactor a zero-power critical assembly, the low neutron flux does not allow the stack irradiation of the monitor foils at the same time of dosimetry foils (Kostal et al., 2020).

## 2 VR-1 reactor

The VR-1 research reactor is a light-water, zero-power pool-type reactor operated by the Czech Technical University in Prague. The core consists of tubular fuel assemblies of IRT-4 type enriched to 19.75 wt. % of $^{235}$U, and contains several dry vertical channels with different diameters up to 90 mm and one radial channel with diameter of 250 mm. The experiment was performed in two different arrangements with the core configuration C13 (see Figure 1) and with the core configuration C12-B (see Figure 2). In both cases the target stacks were placed in the center of a ⌀ 25 mm channel located in the center of the fuel assembly positioned close to the radial channel of the reactor. Due to the reactor core compactness ~42 × 35 × 60 cm and maximum allowable power of approximately 600 W, relatively high fast neutron flux of the order ~1.3·10$^{10}$ n·cm$^{-2}$·s$^{-1}$ can be reached.

Criticality of the reactor during irradiation was ensured by the positions of the control rods. During the first experiment (see Figure 1), they were further away from the target nr. 2. In the second experiment (Figure 2), they were relatively close to target (7 cm above foil set nr. 2). Therefore, the actual spectra in various target arrangements need to be compared. The calculations confirm that the neutron spectra in all arrangements are undistinguishable.

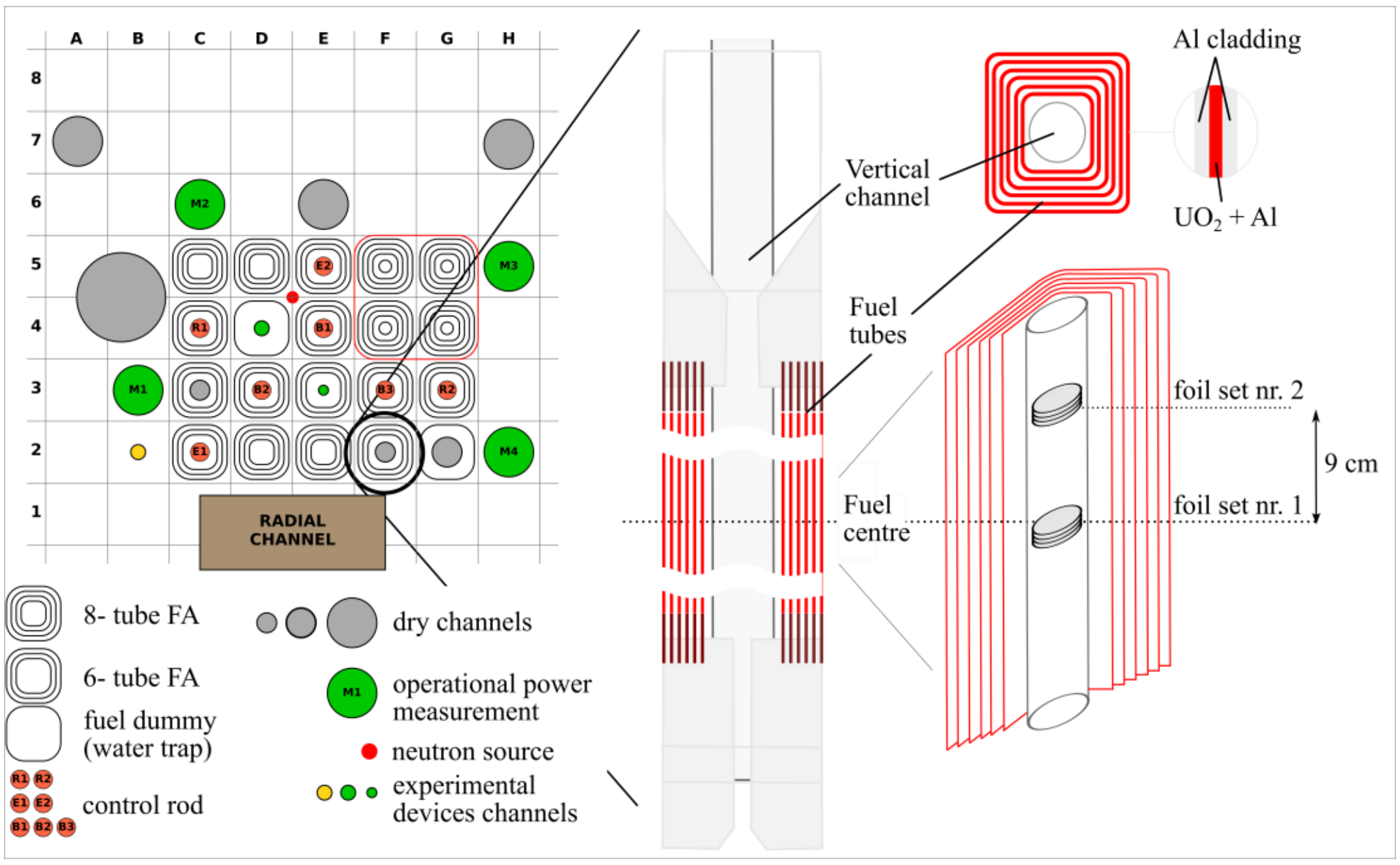

Figure 1: View on VR-1 reactor core (C13) used in the first experiment, together with position of target arrangements.

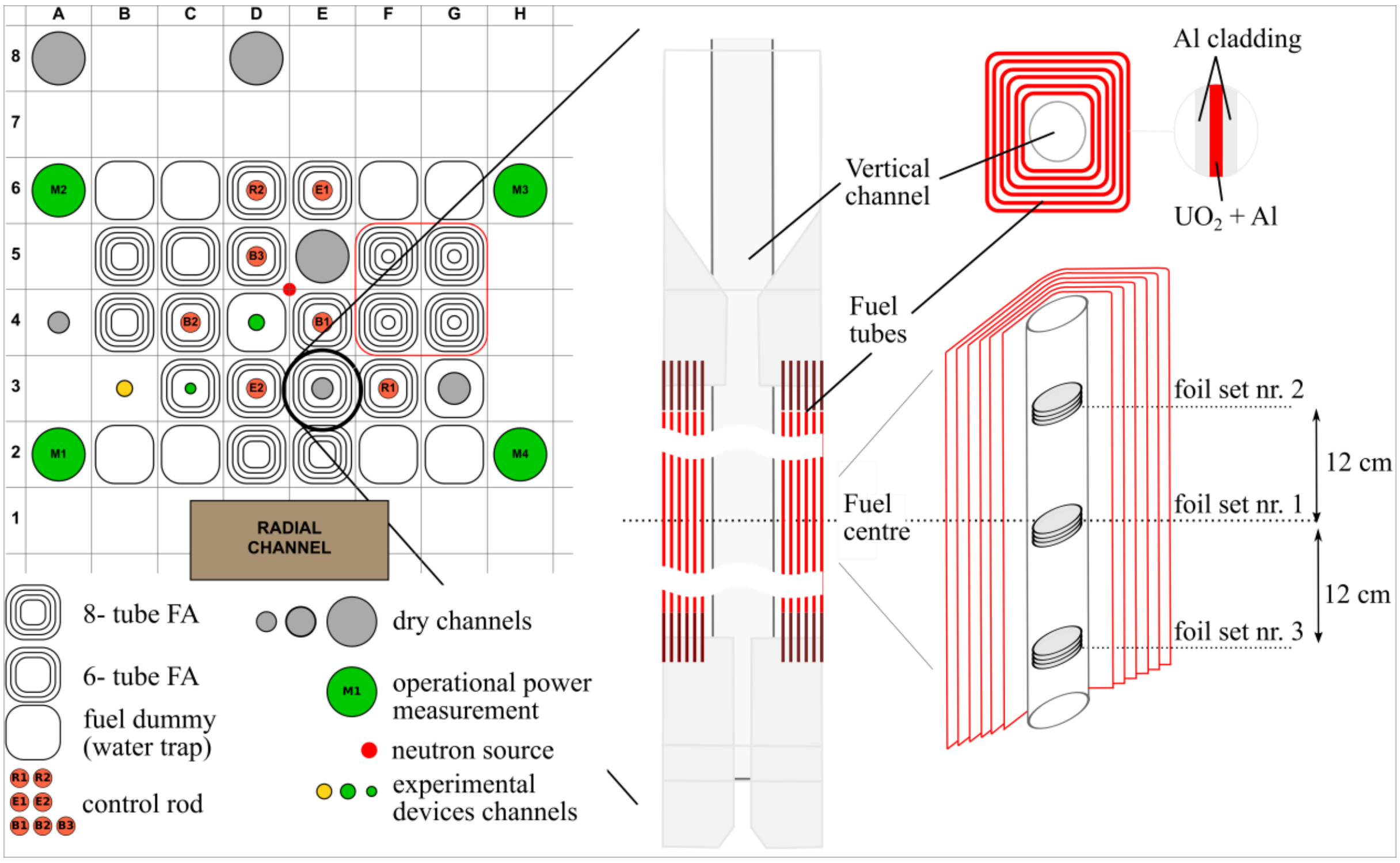

Figure 2: View of the VR-1 reactor core (C12-B) used in the second experiment, together with the position of activated foil stacks.

## 3 Experimental and calculation methods

### 3.1 Irradiation setup

Two independent experiments were realized, at which various foils were irradiated. Both used cores were composed of the same fuel elements, therefore, the neutron spectra in dry cavity in the center of fuel are generally identical in both cases (see C-13 core in Figure 3, C-12B core in Figure 4). However, some flux differences are observed, that could be a result of different location of the dosimeter stack and the surrounding environment in the reactor. Details of the neutron spectra calculations and their properties are discussed in Section 3.3.

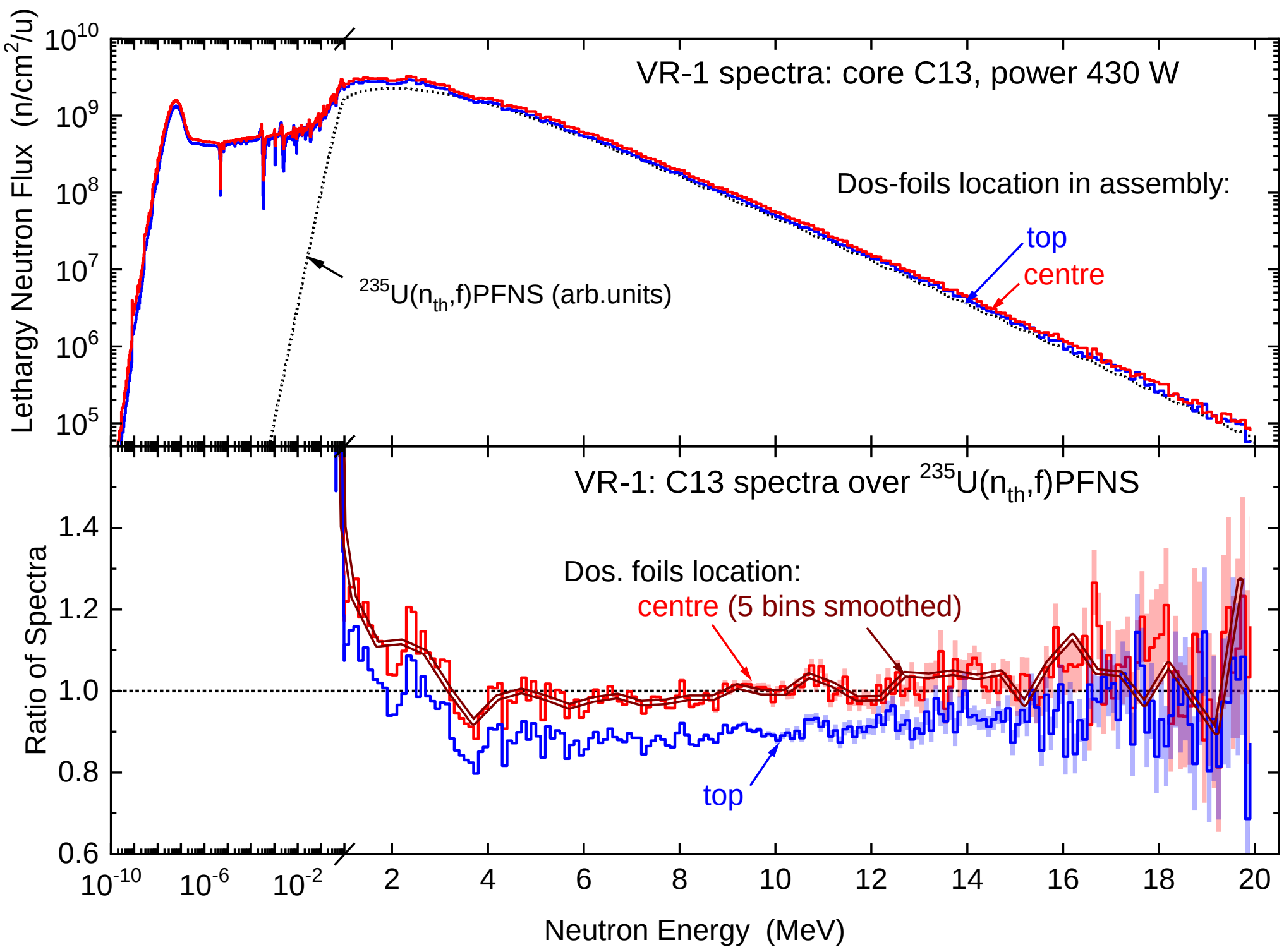

Figure 3: Calculated neutron spectra in the vertical channel of the VR-1 reactor with core configuration C13. The bottom half of Figures show the ratio of the corresponding VR-1 spectra over ENDF/B-VIII.0 $^{235}$U(n$_{th}$,f) PFNS (double curve depicts the ratio smoothed over 5 bins • 0.1 MeV = 0.5 MeV).

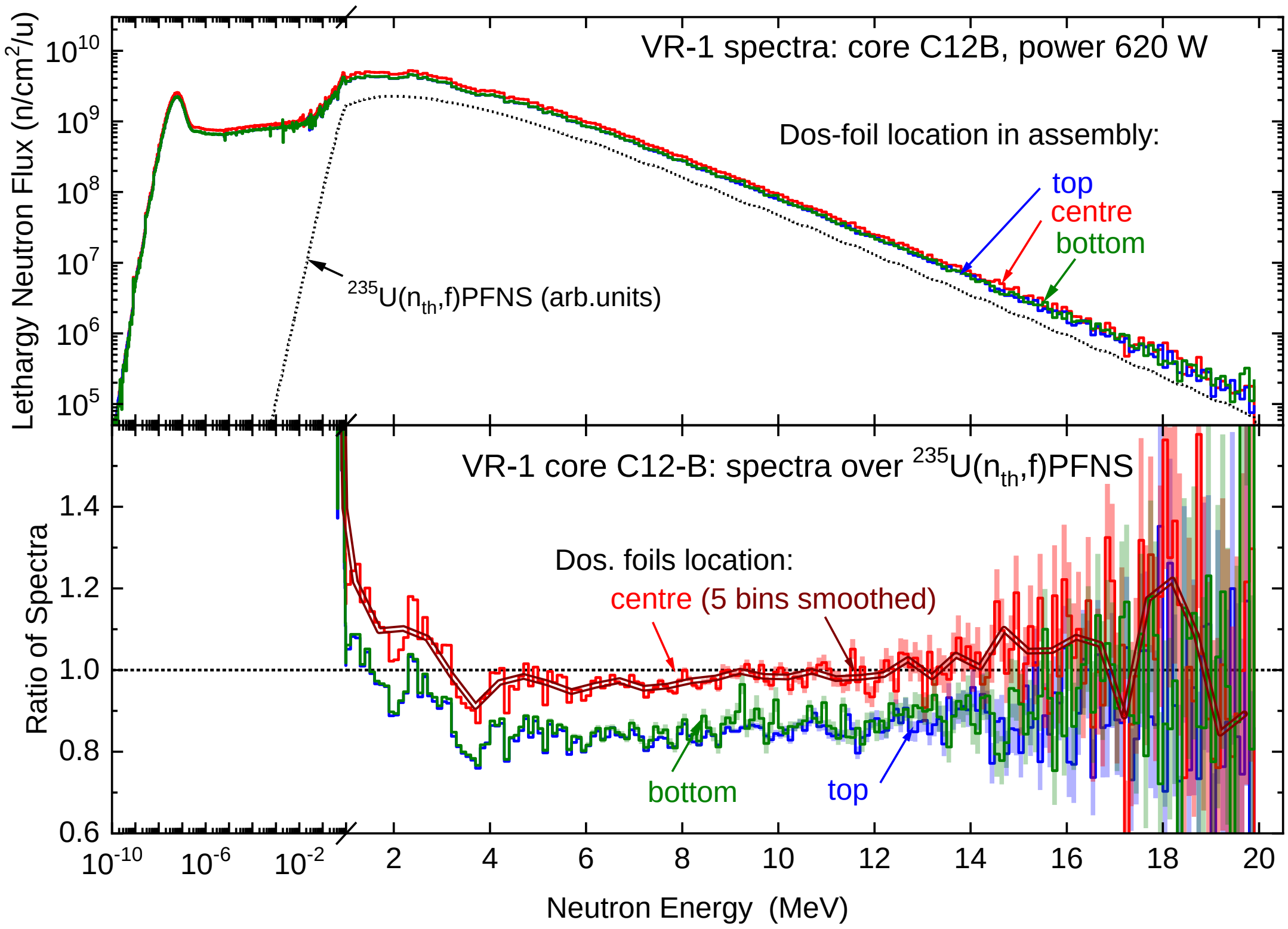

Figure 4: Calculated neutron spectra in the vertical channel of the VR-1 reactor with core configuration C12-B, which corresponds to the separate experimental arrangement. The bottom half of Figures show the ratio of the corresponding VR-1 spectra over ENDF/B-VIII.0 $^{235}$U(n$_{th}$,f) PFNS (double curve depicts the ratio smoothed over 5 bins • 0.1 MeV = 0.5 MeV).

As the reactor power and related neutron flux are enough for good activation of studied foils characterized by high-threshold dosimetry reactions, small activation foils placed in the stack can be used. Thin Ni (0.1mm) and Al (0.25 mm) monitoring foils were placed between the activation foils for SACS measurements. In total, 5 stacks in 2 irradiation experiments were irradiated. The estimated fast neutron flux above 10 MeV in the target arrangement was $7.9 \cdot 10^6$ $cm^{-2} \cdot s^{-1}$ in the first experiment, and $1.3 \cdot 10^7$ $cm^{-2} \cdot s^{-1}$ in the second irradiation. During the first experiment, in the C13 core, 2 stacks were irradiated, one in the center of the fuel and the second at the position 9 cm above the center. During the second experiment, in the C12-B core, 3 stacks were irradiated, one at the center and the other two 12 cm below and above the core center. The stack in the center was composed of thin foils (see Table 1). The thicker foils (Fe, Cu, $CF_2$) were placed in both upper and lower stack to ensure that they will have a negligible impact on the neutron field in the center target position. The equivalent thermal power during the first experiment was 430 W and 620 W during the second experiment.

Due to the piecewise continuous and nearly constant character of power evolution during experiment, the precise $A/A_{sat}$ from the Equation (1) was used.

$$\frac{A(\overline{P})}{A_{Sat}(\overline{P})} = \sum_i \frac{P_i}{\overline{P}} \times \left(1 - e^{-\lambda . T_i}\right) \times e^{-\lambda . T_i^{End}} \qquad (1)$$

Here $A/A_{sat}$ is the ratio of the activity to saturated activity, $\frac{P_i}{\overline{P}}$ is the relative power in i-th interval of the irradiation period, $T_i$ is the irradiation time in i-th interval of the irradiation period and $T_i^{End}$ is the time from the end of the i-th irradiation interval to the end of irradiation period.

### 3.2 Gamma spectrometry

Experimental reaction rates (Eq. 1) were determined from the measured activity of the dosimetry foils used. These quantities were derived from Net Peaks Areas determined employing HPGe gamma spectrometry. Dosimetry as well as monitoring foils were measured separately using a well-charcaterized HPGe spectrometric system in the Research Center Rez (see Kostal et al., 2018b). The thin foils, namely Mg, Fe, $^{54}Fe$, Mo, and Ni, Al monitors were fixed in the plastic EG-3 type holder for ensuring the measuring geometry. The foils were placed on the top of a coaxial HPGe detector or in case of Ni detector 2 cm over its cap. To minimize the background signal, the detector was placed in a lead shielding with a thin inner copper lining and rubber coating. The remaining background with no sample was measured and subsequently subtracted from the sample spectra. Genie 2000 software (Canberra) was used for the spectra evaluation.

The detector energy calibration was performed before the experiment using standard point sources $^{60}Co$, $^{88}Y$, $^{133}B$, $^{137}Cs$, $^{152}Eu$ and $^{241}Am$; energy uncertainty less than 1.0 keV was achieved throughout the used energy range.

The efficiency curve and appropriate Coincidence Summing Factors (Tomarchio et al., 2009 ) were determined using the MCNP6 code and validated mathematical model of HPGe. The model was compiled employing the actual dimensions of HPGe components measured from a precise radiogram supplied by Czech Metrological Institute (Dryak et al., 2006). The insensitive layer thickness was determined experimentally, based on variation in attenuation of $^{241}$Am beam at various incident angles (Boson et al., 2008). The validation was performed for various measuring geometries: point source on detector cap, point source 10 cm from cap, and Marinelli beaker. In the worst case, point source on cap, the discrepancy between calculation and experiment in relevant gamma energy region is below 1.8%. In other cases, namely 10 cm from cap and Marinelli beaker, representing large volume source, the discrepancy was below 1%. Then, the reaction rates are evaluated by means of the equation 2.

$$q(\overline{\mathrm{P}}) = \left(\frac{A(\overline{\mathrm{P}})}{A_{\mathrm{Sat}}(\overline{\mathrm{P}})}\right)^{-1} \times \frac{NPA}{T_{Live}} \times \frac{1}{\varepsilon \times \eta \times N} \times \frac{\lambda \times T_{\mathrm{Re}\,al}}{\left(1-e^{-\lambda.\mathrm{T}_{\mathrm{Re}\,al}}\right)} \times \frac{1}{e^{-\lambda.\Delta\mathrm{T}}} \times \frac{1}{TSCF} \tag{2}$$

Where:

$q(\overline{\mathrm{P}})$; is the reaction rate of activation during power density $\overline{\mathrm{P}}$ (power in the first day of the irradiation experiment);

$\mathrm{T}_{Live}$; is time of measurement by HPGe, corrected to detector dead time;

$\mathrm{T}_{\mathrm{Re}\,al}$; is time of measurement by HPGe, corrected to detector dead time;

$\Delta\mathrm{T}$; is the time between the end of irradiation and the start of HPGe measurement;

$\lambda$; is decay constant of studied isotope;

$\varepsilon$; is the gamma branching ratio;

$\eta$; is the detector efficiency (the result of MCNP6 calculation);

$N$; is the number of target isotope nuclei;

The uncertainty of count rates, being lower than 1%, includes the following main components: gross peak area, Compton continuum area, background area and propagation of uncertainties in the energy and peak shape calibrations. The detector efficiency uncertainty was determined from the difference between the experimentally determined efficiency and the efficiency determined with a precise mathematical model and is about 1.9%.

Besides above stated uncertainties, there are also other stochastic uncertainties: the radionuclide half-time value or branching ratios. However, these uncertainties are negligible in comparison with the count rate uncertainties.

Sometimes, a question arises concerning the multi-material activation detector. Earlier, it has been shown that evaluation of more reaction from one material Tripathy et al., 2007 or nuclei Kostal et al., 2019 is efficient and decreases the uncertainty, especially, in spectra deconvolution method. This was the reason why the possibility of using the stainless steel as a dosimeter was also tested. The measured gamma spectrum of irradiated 08CH18N10-T stainless steel is plotted in Figure 5. The most significant peak is from $^{51}$Cr obtained by activation of $^{50}$Cr by thermal neutrons. The measurement confirms good usability of $^{58}$Ni(n,p)$^{58}$Co, $^{54}$Fe(n,p)$^{54}$Mn and $^{59}$Co(n,$\gamma$)$^{60}$Co reactions in retrospective dosimetry Greenwood et al., 2007, Ilieva et al., 2009 because all peaks are well distinguishable.

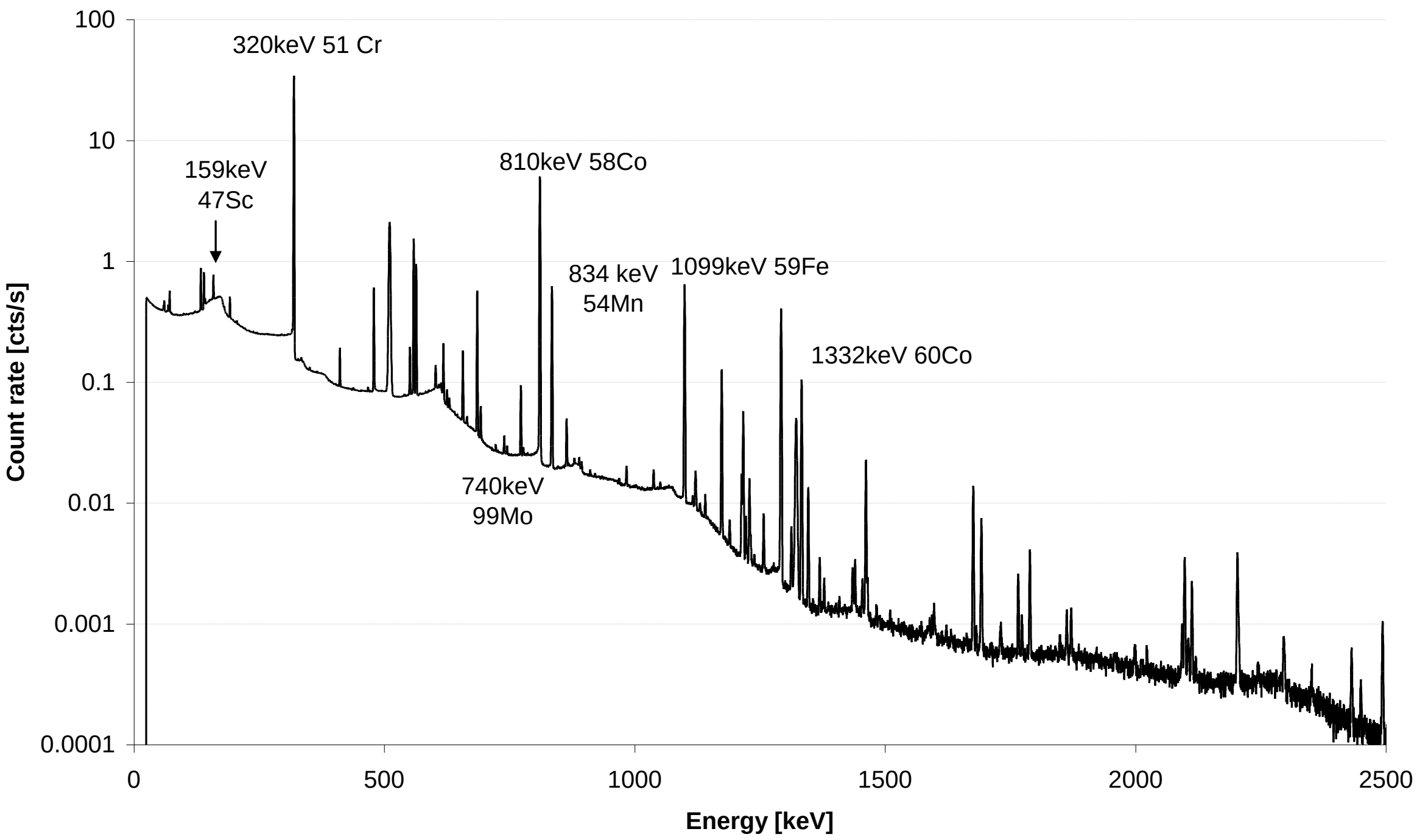

Figure 5: HPGe gamma spectrum of irradiated stainless-steel sample, used as possible flux monitor, 2 days after irradiation.

Table 1.: Summary of used monitors and detectors.

| Reaction | Peak [keV] | Material | Dimensions | Geometry | Efficiency of detection | CSCF |
|---|---|---|---|---|---|---|
| $^{27}Al(n,\alpha)^{24}Na$ | 1368.6 | Al | D=10, th. 0.25mm | EG-3 on cap | 2.951E-2 | 0.863 |
| $^{58}Ni(n,p)^{58}Co$ | 810.8 | Ni | D=10, th. 0.1 | EG-3 on cap | 4.549E-2 | 0.937 |
| $^{56}Fe(n,p)^{56}Mn$ | 846.8 | Fe nat. | D=18, th. 0.1mm | EG-3 on cap | 4.314E-2 | 0.940 |
| | 1810.7 | | | | 2.287E-2 | 0.818 |
| $^{54}Fe(n,p)^{54}Mn$ | 834.8 | Fe nat. | D=18, th. 0.1mm | EG-3 on cap | 4.364E-2 | 1.000 |
| $^{24}Mg(n,p)^{24}Na$ | 1368.6 | Mg | D=18, th. 0.1mm | EG-3 on cap | 2.918E-2 | 0.864 |
| $^{92}Mo(n,p)^{92m}Nb$ | 934.4 | Mo | D=18, th. 0.1mm | EG-3 on cap | 3.980E-2 | 1.000 |
| $^{19}F(n,2n)^{18}F$ | 511.0 | CF2 | D=18, th. 4mm | on cap | 6.284E-2 | 1.000 |
| $^{93}Nb(n,2n)^{92m}Nb$ | 934.4 | Nb | D=18, th. 1mm | on cap | 4.135E-2 | 1.000 |
| $^{58}Ni(n,2n)^{57}Ni$ | 1377.6 | Ni | D=18, th. 1mm | 2.56cm from cap | 9.535E-3 | 0.925 |
| $^{60}Ni(n,p)^{60}Co$ | 1173.0 | | | | 1.084E-2 | 0.941 |
| | 1332.5 | | | | 9.773E-3 | 0.936 |
| $^{58}Ni(n,x)^{57}Co$ | 122.0 | | | | 5.050E-2 | 1.000 |
| $^{54}Fe(n,p)^{54}Mn$ | 834.8 | Fe (98.2%) | D=18, th. 2mm | on cap | 4.273E-2 | 1.000 |
| $^{54}Fe(n,p)^{54}Mn$ | 834.8 | 54Fe (99.5%) | D=15.8, th. 0.1mm | EG-3 on cap | 4.394E-2 | 1.000 |
| $^{54}Fe(n,\alpha)^{51}Cr$ | 320.0 | 54Fe (99.5%) | D=15.8, th. 0.1mm | EG-3 on cap | 1.016E-1 | 1.000 |
| $^{63}Cu(n,\alpha)^{60}Co$ | 1173.0 | Cu (99.6%) | D=18, th. 2mm | on cap | 3.249E-2 | 0.826 |
| | 1332.5 | | | | 2.934E-2 | 0.820 |
| $^{55}Mn(n,2n)^{54}Mn$ | 834.8 | Mn (wax | D=13 th. 3.5mm | on cap | 4.193E-2 | 1.000 |
| | | bounded) | | | | |
| $^{55}$Mn(n,2n)$^{54}$Mn | 834.8 | Mn flake | 10 x 10 x 1.2mm | on cap | 4.594E-2 | 1.000 |
| $^{89}$Y(n,2n)$^{88}$Y | 898.0 | Y | D=18, th. 1.27mm | on cap | 4.277E-2 | 0.840 |
| | 1836.1 | | | | 2.367E-2 | 0.817 |
| $^{47}$Ti(n,p)$^{47}$Sc | 159.4 | Ti | 10 x 10 x 0.25 mm | on cap | 1.819E-1 | 1.000 |
| $^{46}$Ti(n,p)$^{46}$Sc | 889.3 | | | | 4.589E-2 | 0.816 |
| $^{48}$Ti(n,p)$^{48}$Sc | 983.5 | | | | 4.225E-2 | 0.640 |
| | 1037.5 | | | | 4.049E-2 | 0.633 |

Table 2.: Summary of measured reaction rates and monitoring reactions rates

| | 1$^{st}$ experiment in C13 | | 2$^{nd}$ experiment in C12-B | |
|---|---|---|---|---|
| Reaction | RR [$s^{-1}$] | Rel. unc. | RR [$s^{-1}$] | Rel. unc. |
| $^{47}$Ti(n,p) | 1.013E-16 | 2.3% | 1.726E-16 | 2.8% |
| $^{54}$Fe(n,p) | 4.502E-16 | 2.5% | 7.499E-16 | 2.2% |
| $^{92}$Mo(n,p)$^{92m}$Nb | | | 6.356E-17 | 2.1% |
| $^{46}$Ti(n,p) | 5.978E-17 | 2.8% | 1.012E-16 | 3.0% |
| $^{60}$Ni(n,p) | | | 1.865E-17 | 4.6% |
| $^{63}$Cu(n,α) | 2.905E-18 | 3.4% | 4.990E-18 | 2.7% |
| $^{54}$Fe(n,α) | 4.510E-18 | 3.0% | 7.707E-18 | 2.9% |
| $^{56}$Fe(n,p) | | | 9.796E-18 | 2.0% |
| $^{48}$Ti(n,p) | 1.652E-18 | 2.4% | 2.776E-18 | 2.9% |
| $^{24}$Mg(n,p) | | | 1.327E-17 | 2.1% |
| $^{197}$Au(n,2n) | 1.754E-17 | 3.3% | | |
| $^{93}$Nb(n,2n)$^{92m}$Nb | | | 4.074E-18 | 2.2% |
| $^{55}$Mn(n,2n) | 1.298E-18 | 2.9% | | |
| $^{89}$Y(n,2n) | 8.987E-19 | 2.2% | | |
| $^{19}$F(n,2n) | | | 7.323E-20 | 3.1% |
| $^{58}$Ni(n,2n) | 2.104E-20 | 14.1% | 3.740E-20 | 3.2% |
| $^{58}$Ni(n,x)$^{57}$Co | 1.366E-18 | 9.2% | 2.242E-18 | 5.6% |
| $^{58}$Ni(n,p) | 5.987E-16 | 3.0% | 1.024E-15 | 2.2% |
| $^{27}$Al(n,α) | 3.845E-18 | 3.1% | 6.419E-18 | 2.2% |

### 3.3 Calculation methods and VR-1 spectra at the irradiation point.

The simulations of both neutron and photon transport were performed in criticality calculations using the MCNP6 Monte Carlo code (Werner et al., 2017), ENDF/B-VIII.0 (Brown et al., 2018) data library and detailed MCNP model of the VR-1 reactor. Simulation of the γ-ray transport from the sample to the HPGe detector was performed with a gamma transport model of the detector.

The MCNP model of the VR-1 reactor has been validated against experiments for criticality predictions (Huml et al., 2013), neutron spectrum measurements (Kostal et al., 2018, Losa et al., 2020), reaction rates determination (Rataj et al., 2014) or for kinetics parameters (Bily et al., 2019). The calculation of neutron energy spectra in radial channel were validated by Kostal et al., 2018.

The neutron energy spectrum has been calculated using the track length estimate of the cell flux (F4 type card) in 640 group structure in the cylindrical volume with diameter of 1.8 cm and thickness of 0.5 cm positioned at corresponding target locations. The variance reduction method based on the superimposed mesh weight window generation (Werner et al., 2017) was employed to reduce the Monte-Carlo statistical uncertainty for neutron energies up to 16 MeV.

The calculated VR-1 spectra for both the experimental set-up configurations and the irradiation points are shown in Figure 3 and Figure 4 together with the prompt fission neutron spectrum (PFNS) from thermal fission of $^{235}$U. Their comparison indicates large differences in the thermal and epithermal neutron energy domains, but rather similar shapes above 3-5 MeV. For the more detailed analysis of this energy range the ratio of VR-1 spectra over the $^{235}U(n_{th},f)$PFNS are plotted together with calculation uncertainties in the bottom half of Figure 3 and Figure 4. It is seen that the energy shape of ratio is similar for all irradiation locations and tends to increase by ≈ 5-8% when secondary neutron energy varies from 3-5 MeV to 14-16 MeV.

To get more certain information about the high energy part of the VR-1 spectra the MCNP calculated spectra for three target locations (top, middle and bottom) in the C12-B core configuration and for two locations (top, middle) in the C13 experiment were statistically averaged. The resultant ratio with statistical uncertainties are shown in Figure 6. For the further reduction of the statistical uncertainties the VR-1/PFNS ratio was smoothed over 5 energy bins (i.e. 0.5 MeV interval). Now we clearly see the energy oscillations between 2 and 16 MeV and difference between energy slopes of the VR-1 and $^{235}U(n_{th},f)$ fission spectra.

To understand the origin of such differences in spectra, the transmission of the tubular fuel assembly element (enriched $UO_2$ fuel in Al claddings surrounded by water) was calculated analytically taking the total cross sections from ENDF/B-VIII.0. The optimal agreement was achieved for the effective thickness 1.0 cm for fuel 4.5 cm for water (which sum is of the same order as reactor fuel lattice size ≈ 7.15 cm). Comparing the MCNP and analytically calculated spectra ratios in Figure 6, we conclude that fine structure in the VR-1 spectrum up to ≈ 18 MeV is determined by the energy fluctuating total cross section for $^{16}$O. Whereas the slightly different overall energy trends of VR-1 and pure $^{235}U(n_{th},f)$ spectra are most probably caused by the $^{1}$H(n,tot) cross section which decreases by a factor 2-3 in the considered energy range.

It is essential to state that the high energy part of VR-1 spectrum above 12 MeV is a bit harder than $^{235}U(n_{th},f)$ PFNS. For the highest threshold dosimetry reactions with $E_{50\%}$ between 10 and 15 MeV it means that SACS measured in VR-1 spectrum have to be slightly increased by (3-5)%, if they are transformed in SACS corresponding to the $^{235}U(n_{th},f)$ PFNS field and were normalized to the monitor reactions having $E_{50\%}$ ≈ 6-7 MeV. Note that the use of high-energy monitor reactions (e.g., Au-197(n,2n)) may minimize the small bias at energies above 12 MeV.

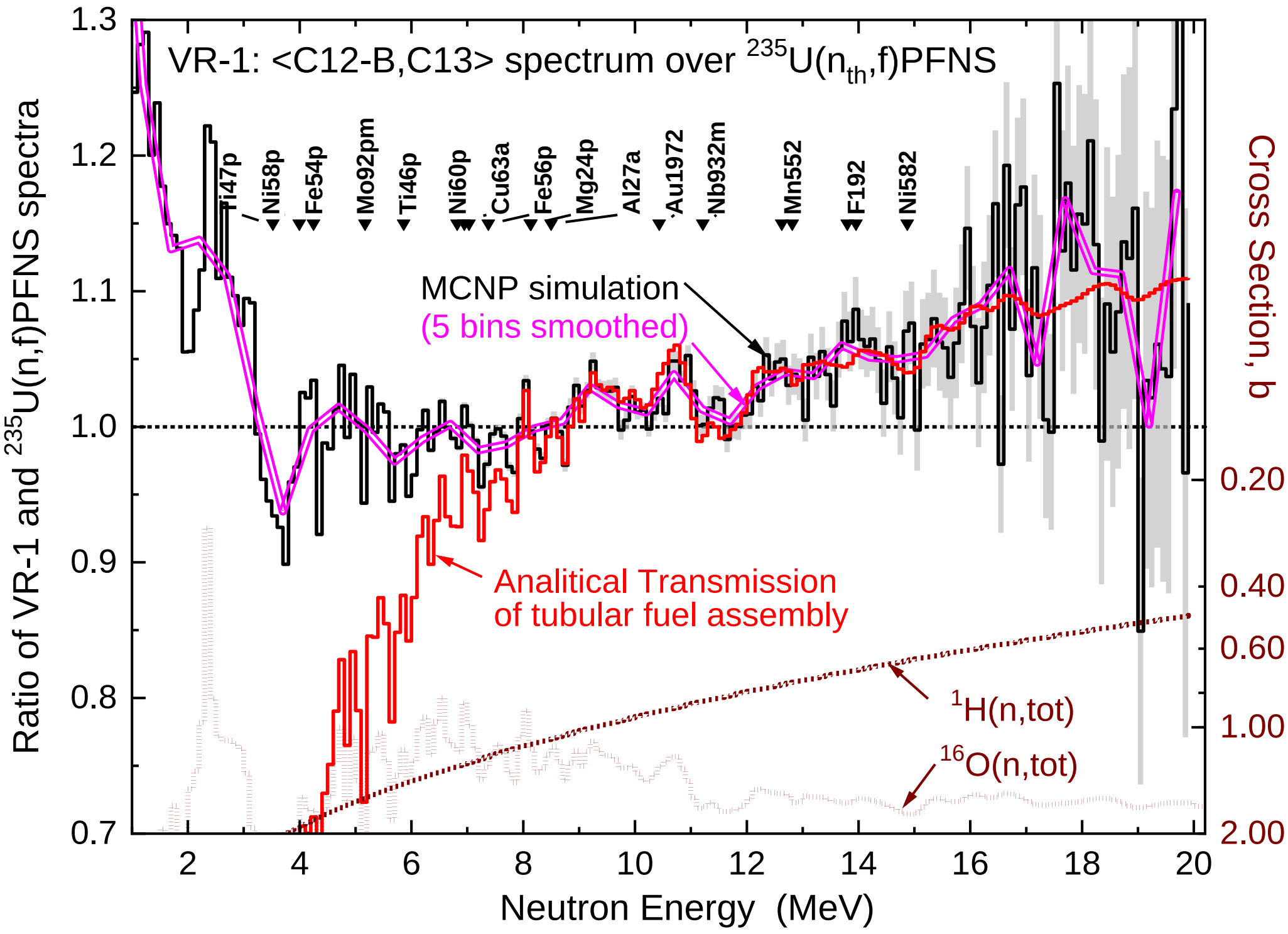

Figure 6: Ratio of the MCNP calculated VR-1 spectrum (averaged for C13-B and C13 cores and 5 sample locations) over $^{235}U(n_{th},f)$ PFNS: MCNP simulation (black curve) and smoothed over 5 energy bins (double pink). Analytically calculated transmission of the fuel assembly cell unit is a red curve. Right hand axis (note downward direction) shows the total neutron cross sections for neutron interaction with $^{1}H$ and $^{16}O$. The $E_{50\%}$ energies for the measured reactions are indicated by triangles.

## 4 Results

The calculation results are standardly normalized per 1 core neutron, while the experiments depend on the used reactor power. Thus for comparison, the experimental reaction rates were also normalized per 1 core neutron using the experimental reaction rates of the monitoring reactions (see Table 2). The scaling factors, see Table 3, are obtained as the ratio between experimental and calculated monitoring reaction rates representing the core neutron emission density during experiment. Based on known fission distribution, the power distribution can be determined as well.

The comparison between calculated and measured reaction rates is presented in Table 6.

Table 3.: Summary of parameters and scaling factors.

| Parameter | 1st experiment (core C13) | 2nd experiment (core C12-B) | Units |
|---|---|---|---|
| Scaling factor | 3.28E+13 | 4.70E+13 | n/s |
| Reactor Power | 434.3 | 622.0 | W |
| Flux > 6 MeV (center) | 1.431E+8 | 1.624E+8 | $[cm^{-2}\cdot s^{-1}]$ |
| Calculated $^{58}Ni(n,p)$ rate | 1.84E-29 | 2.13E-29 | $[s^{-1}]$ |
| Calculated $^{27}Al(n,\alpha)$ rate | 1.16E-31 | 1.40E-31 | $[s^{-1}]$ |

### 4.1 Spectrum averaged cross sections

With knowledge of the neutron flux at the target, reaction rates can be used to determine the spectrum averaged cross section. Two possible methodologies of spectral weighted cross sections evaluation were tested.

In the first methodology, the fluxes in target are determined using calculations. The scaling factor needed for absolute flux is determined from monitors positioned in well-defined positions. This approach is generally applicable but requires a well-validated mathematical model of the reactor core.

The second methodology assumes that neutron flux in monitor and dosimetry foils is identical. To comply with this assumption, very thin monitoring foils were placed between various dosimeter foils. The assumption is met, since the measured reaction rates of the monitoring reactions $^{58}$Ni(n,p) and $^{27}$Al(n,α) were the same in all monitors distributed across the stack target assembly.

#### 4.1.1 Evaluation using calculated fluxes

It has been shown by direct stilbene measurements (Kostal et al., 2018) that in the radial channel the neutron spectrum of VR-1 above 6 MeV is undistinguishable from $^{235}$U(n$_{th}$,f) PFNS. It means that if the threshold reaction with threshold over 6 MeV is measured, the interacting part of neutron spectrum has the same shape as $^{235}$U(n$_{th}$,f) PFNS. As we see from the previous section, a careful study showed differences of 3-5% for the highest energies above 12 MeV. The mathematical model of the VR-1 allows to determine the fluxes in targets due to the knowledge of calculated and measured reaction rates of monitoring reactions (see Table 2). Based on the comparison between calculated and experimental reaction rates (RR) the scaling factor used in evaluations (Kostal et al., 2018b) was determined. The comparison of spectra and energy interval used in normalization is plotted in Figure 7.

In the evaluated ENDF/B-VIII.0 $^{235}$U(n$_{th}$,f) PFNS, about 2.566% of total emitted neutrons have energy above 6 MeV. In the VR-1 spectrum only 0.777% of neutrons have energy above 6 MeV. Thus, for determination of the SACS, the flux which is used for normalization of calculated RR must be divided by a factor 3.303 which reflects that a large amount of thermal and epithermal neutrons is added to the part of "true PFNS". The resulted spectrum averaged cross sections (SACS) are listed in Table 4.

$$\overline{\sigma}^{\mathrm{exp.}} = \frac{1}{K} \times \frac{q\left(\overline{\mathrm{P}}\right)}{\int_{\mathrm{E}} \varphi(\mathrm{E}) \times d\mathrm{E}} \times \mathrm{R} \qquad (3)$$

Where:

$K$; is the scaling factor based on absolute flux density, neutron emission per second (see (3));

$q\left(\overline{\mathrm{P}}\right)$; is the experimentally measured reaction rate (see (2));

$\varphi(\mathrm{E})$; is the calculated neutron spectrum normalized per 1 neutron in core;

$\mathrm{R}$; is the ratio between share of neutrons with energy above 6 MeV in $^{235}$U(n$_{th}$,f) PFNS (2.566% in ENDF/B-VIII.0) and VR-1 spectra (0.777%) being 3.303.

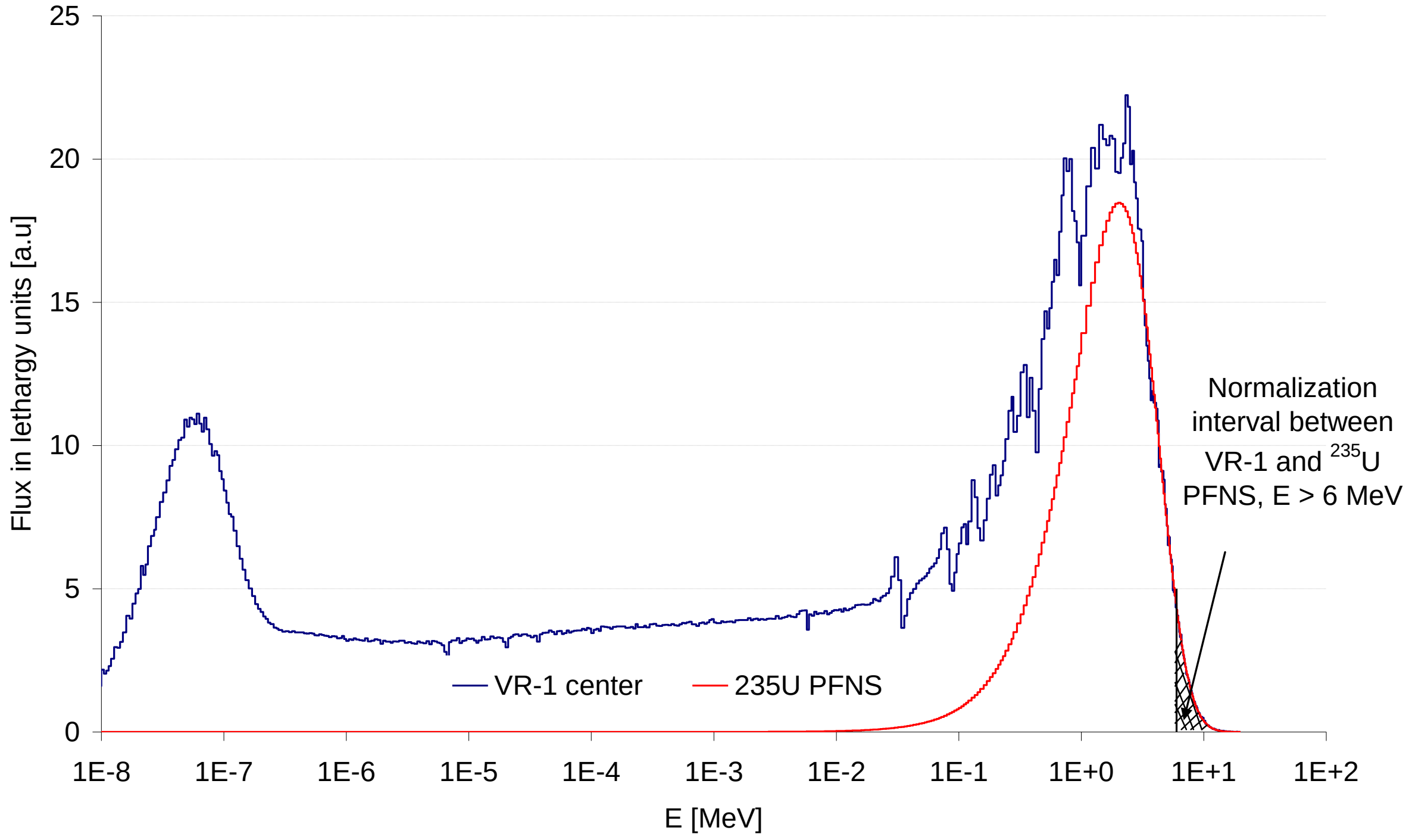

Figure 7.: Graphical interpretation of similarity in VR-1 spectra and $^{235}U(n_{th},f)$ PFNS and following flux normalization.

Table 4.: SACS in ENDF/B-VIII.0 $^{235}U(n_{th},f)$ PFNS evaluated using the calculated neutron flux approach.

| Reaction | Mean [mb] | Unc |
|---|---|---|
| $^{47}Ti(n,p)$ | 18.40 | 3.8% |
| $^{54}Fe(n,p)$ | 80.83 | 3.7% |
| $^{92}Mo(n,p)^{92m}Nb$ | 6.821 | 3.2% |
| $^{46}Ti(n,p)$ | 10.82 | 4.1% |
| $^{60}Ni(n,p)$ | 2.001 | 4.3% |
| $^{63}Cu(n,\alpha)$ | 0.5297 | 4.2% |
| $^{54}Fe(n,\alpha)$ | 0.8202 | 4.1% |
| $^{56}Fe(n,p)$ | 1.051 | 3.2% |
| $^{48}Ti(n,p)$ | 0.2979 | 3.9% |
| $^{24}Mg(n,p)$ | 1.424 | 3.2% |
| $^{197}Au(n,2n)$ | 3.162 | 3.7% |
| $^{93}Nb(n,2n)^{92m}Nb$ | 0.4372 | 3.2% |
| $^{55}Mn(n,2n)$ | 0.2340 | 3.5% |
| $^{89}Y(n,2n)$ | 0.1621 | 3.3% |
| $^{19}F(n,2n)$ | 0.007858 | 3.6% |
| $^{58}Ni(n,2n)$ | 4.01E-03 | 4.3% |
| $^{58}Ni(n,x)^{57}Co$ | 2.41E-01 | 6.3% |

#### 4.1.2 Evaluation using the normalization to monitor cross sections

All detector foils from target assembly were surrounded by monitoring Al and Ni foils. Thus, one can assume the neutron flux in detector and monitoring foils is identical. This assumption was verified experimentally by measuring the gamma activities of specific monitoring foils. It was observed that the measured reaction rates of monitoring foils placed between detectors are nearly identical, independently of the monitor foil position in the stack. The target foils were placed in the dry channel of diameter 25 mm located in the center of fuel assembly. Due to the proximity of the fission region, the same normalization approach can be used as in works Steinnes E., 1970, Arribére et al., 2001 or Maidana et al., 1994. This is supported by corresponding ratio between averaged cross section ratio of $^{58}$Ni(n,p) and $^{27}$Al(n,α) reactions at VR-1 reactor, which is 157.7 in the first experiment, 152.6 in the second experiment, while in the ENDF/B-VIII.0 $^{235}$U($n_{th}$,f) PFNS is 154.4. The final normalization was realized using the $^{27}$Al(n,α) monitor reaction, which has threshold over 6 MeV to minimize the differences between the reported VR-1 spectra and the ENDF/B-VIII.0 $^{235}$U($n_{th}$,f) PFNS. The IRDFF value of 0.7007 mb for the $^{235}$U($n_{th}$,f) PFNS SACS for $^{27}$Al(n,α) was used. Resulting SACS are listed in Table 5.

Results are in good agreement with those obtained by the evaluation using the calculated neutron flux. Therefore, this can be understood as a validation of the presented methodology.

Very interesting result showing the possible use of new dosimetry reaction was obtained for $^{58}$Ni(n,x)$^{57}$Co. Averaging the available data of BRUGGEMAN et al., 1974 0.216 ± 0.005, WÖLFLE et al., 1980 0.240 ± 0.035; HORIBE et al., 1992 0.232 ± 0.005; ZAIDI et al., 1993 0.253 ± 0.015; Arribére et al., 2001 0.275 ± 0.015, Burianova et al., 2019 0.239 ± 0.013 we have obtained the value 0.243 ±0.018 mb which is in very good agreement with current results being 0.241 mb. This reaction is interesting due to relatively long half-life being 271.74 d, thus can be an excellent monitor for reactor dosimetry. Other specific feature of this reaction is a high threshold, which makes it interesting for characterization of accelerator neutron fields Kostal et al., 2019.

Table 5.: SACS in $^{235}$U($n_{th}$,f)PFNS evaluated using $^{27}$Al(n,α) monitoring cross sections

| Reaction | Mean [mb] | Unc |
|---|---|---|
| $^{47}$Ti(n,p) | 18.47 | 3.8% |
| $^{54}$Fe(n,p) | 81.15 | 3.7% |
| $^{92}$Mo(n,p)$^{92m}$Nb | 6.828 | 3.2% |
| $^{46}$Ti(n,p) | 10.86 | 4.1% |
| $^{60}$Ni(n,p) | 2.003 | 4.3% |
| $^{63}$Cu(n,α) | 0.5320 | 4.2% |
| $^{54}$Fe(n,α) | 0.8231 | 4.1% |
| $^{56}$Fe(n,p) | 1.052 | 3.1% |
| $^{48}$Ti(n,p) | 0.2991 | 3.9% |
| $^{24}$Mg(n,p) | 1.426 | 3.2% |
| $^{197}$Au(n,2n) | 3.184 | 3.7% |
| $^{93}$Nb(n,2n)$^{92m}$Nb | 0.4377 | 3.2% |
| $^{55}$Mn(n,2n) | 0.2356 | 3.5% |
| $^{89}$Y(n,2n) | 0.1631 | 3.2% |
| $^{19}$F(n,2n) | 0.007867 | 3.6% |
| $^{58}$Ni(n,2n) | 4.02E-03 | 4.3% |
| $^{58}$Ni(n,x)$^{57}$Co | 2.41E-01 | 6.3% |

The spectrum average dosimetry cross sections (SACS) with the MCNP simulated neutron spectra shown in Figure 3 and Figure 4 were computed with the help of the code RR_UNC (Trkov et al., 2001). The dosimetry cross sections and their covariances were taken from the IRDFF-II library (Trkov et al., 2020). The C/E ratios for SACS using measured data were normalized to get the monitoring reactions $^{27}$Al(n,α) and $^{58}$Ni(n,p) close to unity. The results are summarized in Table 6 and displayed in Figure 8, for experiment #1 (core C13) and Figure 9 for experiment #2 (core C12-B), correspondingly. It is worth noticing, that SACS uncertainties resultant from the MCNP simulation of the VR-1 spectra are smaller than those propagated from the IRDFF-II cross sections even for high-threshold reactions.

We observe as a rule an agreement within estimated experimental and calculated uncertainties for the most of reactions, except reaction $^{54}$Fe(n,α). For the latter we are suspicious about the foil enrichment provided by supplier. An overestimation 5-10% is observed for the highest threshold (n,2n) reactions with mean energy response between 10 and 15 MeV. Such disagreement, which however is within two sigma uncertainties, may point to slight overestimation of the ENDF/B-VIII.0 $^{235}$U($n_{th}$,f) PFNS between 10 and 15 MeV.

Reaction $^{58}$Ni(n,x)$^{57}$Co is not part of the IRDFF-II library. Its cross section was taken from the ENDF/B-VIII.0 evaluation. Production of $^{57}$Co is possible through several reaction pathways: (n,np + pn), (n,d) and (n,2n)$^{57}$Ni($\beta^+$, $T_{1/2}$ = 35.6 h). The agreement between the ENDF/B-VIII.0 and present measurements (averaged over two experiments C13 and C12-B) turns out to be quite acceptable: C/E = 0.95 ± 0.05.

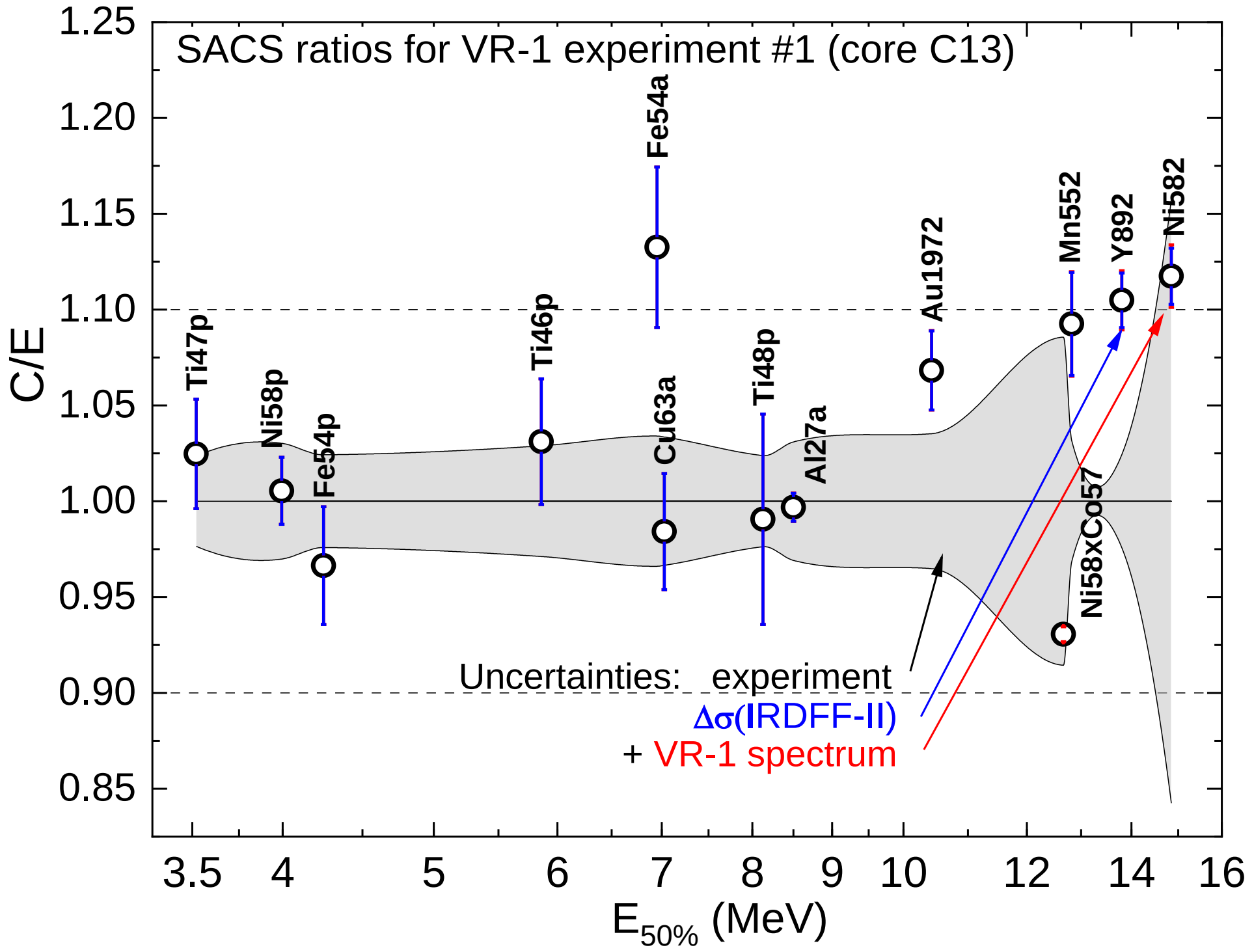

Figure 8: The SACS C/E ratio for the dosimetry foils irradiated in the center of the dry channel of the VR-1 reactor fuel assembly (core C13). The measurement uncertainties are shown by grey corridor, the contribution from IRDFF-II cross sections – blue bars, neutron spectrum – red bars. For the $^{58}$Ni(n,x)$^{57}$Co the C/E computed with ENDF/B-VIII.0 is plotted.

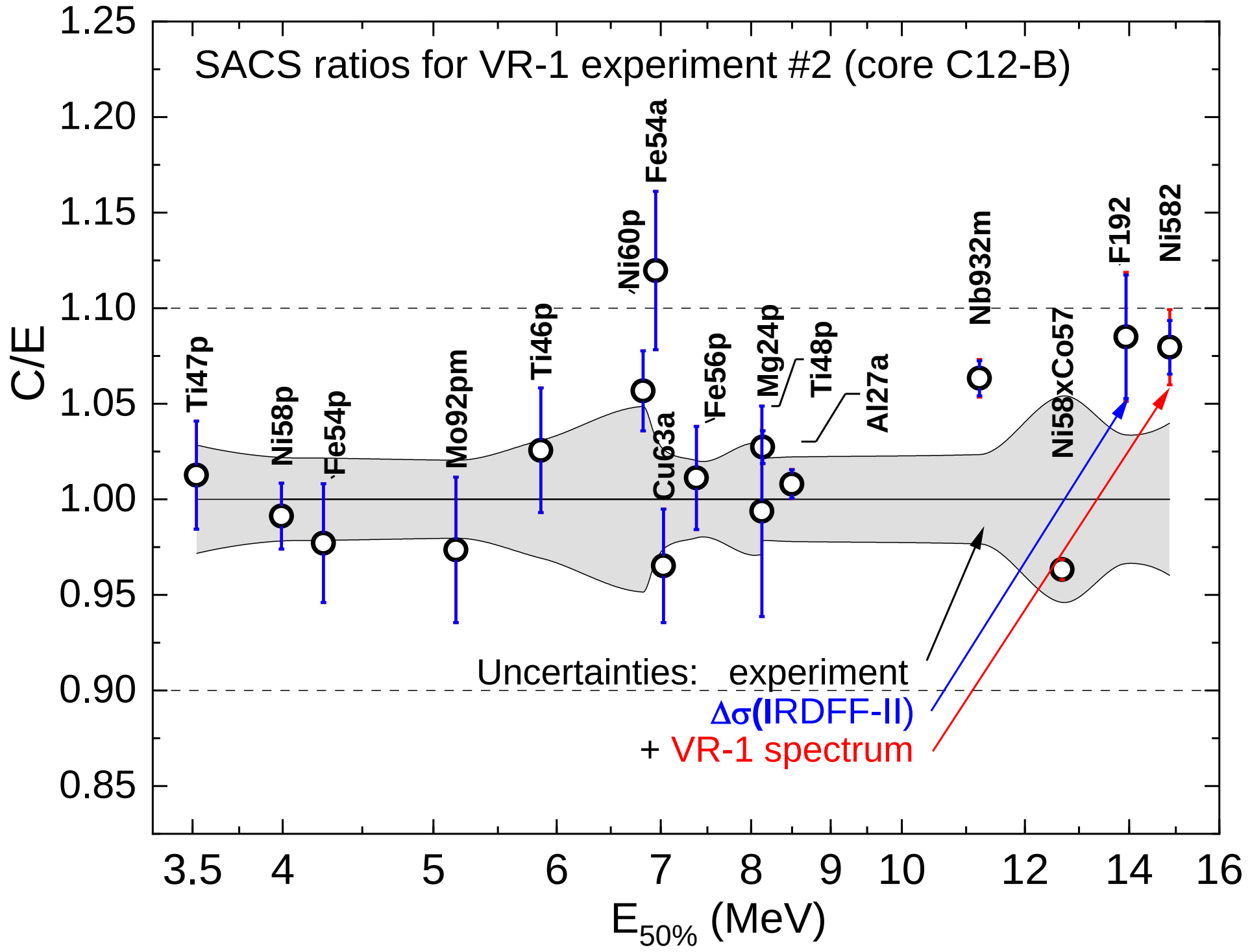

Figure 9: The SACS C/E ratio for the dosimetry foils irradiated in the center of the dry channel of the VR-1 reactor fuel assembly (core C12-B). The measurement uncertainties are shown by grey corridor, the contribution from IRDFF-II cross sections – blue bars, neutron spectrum – red bars. For the $^{58}Ni(n,x)^{57}Co$ the C/E computed with ENDF/B-VIII.0 is plotted.

Table 6.: SACS measured in two VR-1 reactor core configurations s (C13 and C12-B) and C/E ratios computed with cross sections from IRDFF-II except reaction *$^{58}Ni(n,x)^{57}Co$ – from ENDF/B-VIII.0.*

| Reaction | $E_{50\%}$ [MeV] | 1. experiment in C13 | | | | 2. experiment in C12-B | | | |
|---|---|---|---|---|---|---|---|---|---|
| | | SACS [mb] | Unc. % | C/E | ΔC/E [%] | SACS [mb] | Unc. [%] | C/E | ΔC/E [%] |
| $^{47}$Ti(n,p) | 3.521 | 3.088E-03 | 2.3 | 1.025 | 3.6 | 3.672E-03 | 2.8 | 1.013 | 4.0 |
| $^{54}$Fe(n,p) | 4.248 | 1.373E-02 | 2.5 | 0.966 | 4.1 | 1.596E-02 | 2.2 | 0.977 | 3.9 |
| $^{92}$Mo(n,p)$^{92m}$Nb | 5.170 | | | | | 1.352E-03 | 2.1 | 0.974 | 4.4 |
| $^{46}$Ti(n,p) | 5.861 | 1.823E-03 | 2.8 | 1.031 | 4.2 | 2.153E-03 | 3.0 | 1.026 | 4.4 |
| $^{60}$Ni(n,p) | 6.820 | | | | | 3.968E-04 | 4.6 | 1.057 | 5.0 |
| $^{54}$Fe(n,α) | 6.947 | 1.375E-04 | 3.0 | 1.133 | 4.8 | 1.640E-04 | 2.9 | 1.120 | 4.7 |
| $^{63}$Cu(n,α) | 7.029 | 8.857E-05 | 3.4 | 0.984 | 4.6 | 1.062E-04 | 3.1 | 0.965 | 4.1 |
| $^{56}$Fe(n,p) | 7.378 | | | | | 2.084E-04 | 2.0 | 1.011 | 3.3 |
| $^{48}$Ti(n,p) | 8.129 | 5.037E-05 | 2.4 | 0.991 | 6.0 | 5.906E-05 | 2.9 | 0.994 | 6.3 |
| $^{24}$Mg(n,p) | 8.139 | | | | | 2.823E-04 | 2.1 | 1.027 | 2.3 |
| $^{197}$Au(n,2n) | 10.433 | 5.348E-04 | 3.3 | 1.068 | 3.8 | | | | |
| $^{93}$Nb(n,2n)$^{92m}$Nb | 11.214 | | | | | 8.668E-05 | 2.2 | 1.063 | 2.4 |
| $^{55}$Mn(n,2n) | 12.813 | 3.957E-05 | 2.9 | 1.093 | 3.8 | | | | |
| $^{89}$Y(n,2n) | 13.799 | 2.740E-05 | 2.2 | 1.105 | 2.6 | | | | |
| $^{19}$F(n,2n) | 13.955 | | | | | 1.558E-06 | 3.0 | 1.085 | 4.4 |
| $^{58}$Ni(n,2n) | 14.871 | 6.415E-07 | 14.1 | 1.117 | 14.2 | 7.957E-07 | 3.7 | 1.080 | 4.2 |
| *$^{58}Ni(n,x)^{57}Co$* | *12.628* | *4.165E-05* | *9.2* | *0.931* | *9.2* | *4.770E-05* | *5.6* | *0.963* | *5.6* |
| $^{58}$Ni(n,p) | 3.994 | 1.825E-02 | 3.0 | 1.005 | 3.5 | 2.179E-02 | 2.2 | 0.991 | 2.8 |
| $^{27}$Al(n,α) | 8.499 | 1.172E-04 | 3.1 | 0.997 | 3.2 | 1.366E-04 | 2.2 | 1.008 | 2.3 |

## 5 Conclusions

In this work, a large set of spectrum averaged cross section (SACS) has been measured. The results are self-consistent, two different evaluation methods were used, and different experimental conditions were tested. The experiments were realized in experimental channel placed at the center of the reactor core formed by IRT-4M fuel in VR-1 reactor. The used thin foils ensure negligible perturbation on the fast neutron flux. The measurement of the decay γ-rays was realized in extremely well-characterized HPGe spectrometric system in Research Centre Řež. The presence of thin Al and Ni monitor foils interspersed between dosimetry foils allows also the SACS evaluation by both calculated neutron flux and also normalization to the monitor reactions as the variation of the neutron flux across the stack was experimentally verified to be negligible.

The neutron spectra in the dry channels of the thermal research reactor VR-1 were calculated by the Monte Carlo code MCNP employing the neutron fission and transport data from ENDF/B-VIII.0. The statistical uncertainty in the calculated spectra up to 16 MeV was minimized with help of a variance reduction technique based on the weight window generation. Due to this it became possible to reveal the difference of the VR-1 spectrum above 10 MeV from the ENDF/B-VIII.0 $^{235}U(n_{th},f)$ prompt spectrum, and to find an explanation by performing analytical transmission analysis. Thus a larger energy gradient of +10%/10MeV of the VR1 spectrum above 4-5 MeV in comparison with $^{235}U(n_{th},f)$ PFNS is caused by decreasing $^{1}H(n,tot)$ cross section. The statistically significant local oscillations in the VR-1 spectrum between 2 and 17 MeV was shown to stem from the resonance type structure of $^{16}O$(n,tot). In general, we can expect similar spectral differences in reactor systems with water as a moderator and fuel cooling media.

The measured spectrum averaged IRDFF-II cross sections for 16 dosimetry threshold reactions were validated in the VR-1 reactor neutron fields. An agreement within 1-2 experimental uncertainties is observed except for the reaction $^{54}Fe(n,\alpha)$, where the supplier's documented foil enrichment is suspected to be wrong. A 5-10% systematic overestimation observed for the high threshold dosimetry (n,2n) reactions could be mindful indication of the slight overestimation of the ENDF/B-VIII.0 $^{235}U(n_{th},f)$ PFNS between 10 and 15 MeV.

Present results validate the ENDF/B-VIII.0 evaluation for the $^{58}Ni(n,x)^{57}Co$ reaction. The suitable properties of this reaction make it a candidate for the next update of the IRDFF library. This reaction will be beneficial for the reactor dosimetry and characterization of the accelerator high-energy neutron fields.

## 6 Acknowledgements

The presented work was financially supported by the Project CZ.02.1.01/0.0/0.0/15_008/0000293: Sustainable energy (SUSEN) – 2nd phase, realized in the framework of the European Structural and Investment Funds and with use of infrastructures Reactors LVR-15 and LR-0, which were financially supported by the Ministry of Education, Youth and Sports - projects LM2015093 and LM2015074.
The authors would like to thank to VR-1 staff headed by F. Fejt for their effective help during the experiments and for the precise monitoring of the reactor power. The VR-1 operation was supported by LM2018118 project: VR-1 - Support for reactor operation for research activities, which was granted by The Ministry of Education, Youth and Sports of the Czech Republic.